\documentclass{aa}  

\usepackage{graphicx}
\usepackage{soul}
\usepackage{float} 
\usepackage{placeins}
\usepackage{txfonts}
\usepackage[dvipsnames]{xcolor}
\usepackage{verbatim}
\begin{document}

   \title{Multi-instrument STIX microflare study}

  \author{Jonas Saqri\inst{1} \and
      Astrid M. Veronig\inst{1} \and
      Alexander Warmuth\inst{2} \and 
      Ewan C. M. Dickson\inst{1,3} \and
  Andrea Francesco Battaglia\inst{3,4} \and
      Tatiana Podladchikova\inst{5}\and
    Hualin Xiao\inst{3} \and
    Marina Battaglia\inst{3} \and 
    Gordon J. Hurford\inst{3} \and 
    S\"am Krucker\inst{3,6}
     }

   \institute{
   	         Institute of Physics, University of Graz, A-8010 Graz, Austria
		     \and 
		                  Leibniz-Institut f\"ur Astrophysik Potsdam (AIP), An der Sternwarte 16, D-14482 Potsdam, Germany
             \and
             University of Applied Sciences and Arts Northwestern Switzerland, Bahnhofstrasse 6, 5210 Windisch, Switzerland 
             \and
             ETH Z\"urich, R\"amistrasse 101, 8092 Z\"urich, Switzerland 
             \and
             Skolkovo Institute of Science and Technology, Bolshoy Boulevard 30, bld. 1, Moscow 121205, Russia
                         \and
             Space Sciences Laboratory, University of California, 7 Gauss Way, 94720 Berkeley, USA
}
\authorrunning{J. Saqri et~al.}

   \date{}

 
  \abstract
   {During its commissioning phase in 2020, the Spectrometer/Telescope for Imaging X-rays (STIX) on board the Solar Orbiter spacecraft observed 69 microflares. The two largest events from this set (of GOES class B2 and B6), both of which were observed on-disk from the spacecraft as well as from Earth, are analysed in terms of the spatial, temporal and spectral characteristics.}
   {We complement observations from the STIX instrument with EUV imagery from SDO/AIA and GOES soft X-ray data to add imaging and plasma diagnostics over different temperature ranges for a detailed microflare case study in terms of energy release and transport.
   }
   {We use data from GOES for SXR plasma diagnostics and SDO/AIA for high cadence EUV imaging and the reconstruction of differential emission measure (DEM) maps of the thermal flare plasma. The reconstructed DEM profiles are used to study the temporal evolution of thermal flare plasma in the kernels and loops independently. We derive the time evolution of the flare plasma parameters (EM, T) and thermal energy from STIX, GOES and AIA observations, and study STIX spectra to determine the nonthermal emission from accelerated electrons. 
   }
   {Spectral fitting of the STIX data shows clear nonthermal emission for both microflares under study. For both events the plasma temperature and EM derived from STIX, GOES as well as the reconstructed DEM maps differ in absolute values and timing with AIA (sensitive to lower plasma temperatures) lagging behind. The deduced plasma parameters from either method roughly agree with values in the literature for microflares as do the nonthermal fit parameters from STIX. This finding is corroborated by the Neupert effect exhibited between the time derivative of the GOES SXR emission and the STIX HXR profiles. For the B6 event where such an analysis was possible, the nonthermal energy deduced from STIX roughly coincides with the lower estimates of the thermal energy requirement deduced from the SXR and EUV emissions.
   }
   {
   The observed Neupert effects and impulsive/ gradual phases indicate that both events under study are consistent with the standard chromospheric evaporation flare scenario. For the B6 event on June 7th, 2020, this interpretation is further supported by the temporal evolution seen in the DEM profiles of the flare ribbons and loops. For this event we further find that accelerated electrons can roughly account for the required thermal energy. The June 6th, 2020 event demonstrates that STIX can detect nonthermal emission for GOES class B2 events albeit smaller than the background rate level. We demonstrate for the first time how detailed multi-instrument studies of solar flares can be performed with STIX. 
   }
   \maketitle
\section{Introduction}
 Solar flares are generally thought to be a result of the impulsive release of previously stored free magnetic energy by magnetic reconnection (e.g. review by \citealt{priestforbes2002}). This release of energy gives rise to a wide range of solar phenomena that influence the heliosphere and, in some cases even conditions on Earth (e.g. ionization events in the atmosphere and radio disturbances, \citealt{Koskinen2017}). Part of this liberated energy is used for particle acceleration and to heat up the solar plasma (reviews by \citealt{fletcher2011,benz2017}). The heated thermal flare plasma can be imaged with EUV instruments such as the Atmospheric Imaging Assembly (AIA) onboard the Solar Dynamics Observatory (SDO) spacecraft. X-ray photons produced by the thermal bremsstrahlung process of the hot plasma are observed by the Spectrometer/Telescope for Imaging X-rays (STIX, \citealt{stix2012,stix2020V2}) on board Solar Orbiter \citep{mueller2020}. The measured X-ray spectrum further contains a signature of the downwards propagating electrons accelerated in the flare process because they produce nonthermal hard X-ray (HXR) emission via bremsstrahlung when decelerating as they encounter denser chromospheric plasma (\citealt{brown1971,linHudson1976,holman2011}).
 
For many flares, there exists a tendency for the derivative of the observed SXR flux to match the HXR lightcurve in flares which was termed "Neupert effect" by \cite{hudson1991} after Werner Neupert who first reported that time integrated microwave flux and SXR flux often show similar time profiles \citep{neupert1968}. This relation between the thermal and nonthermal flare emissions was confirmed in a number of statistical studies relating the HXR and SXR profiles in flares (e.g. \citealt{dennisZarro1993,Veronig2002}). The derivative of the SXR time profile matches the HXR radiation because the SXR emission is generated by thermal emission from the plasma heated by the flare accelerated electrons and convected into the corona by a process called chromospheric evaporation \citep{veronig2005}.
 
 Flares of GOES class B and smaller, are usually termed microflares \citep{hannah2011}. Like their larger counterparts, they occur in active regions \citep{stoiser2007,christie2008} and typically show an impulsive HXR phase followed by gradual thermal SXR emission. Recently, the Neupert effect was reported for a microflare of GOES class A5.7 from NuSTAR observations \citep{glesener2020}. Based on over 4000 microflares observed with RHESSI, the power law index of the nonthermal electrons $\delta$ lies within the range of 4-10 for 90\% of the events. The thermal emission for 90\% of over 9100 analysed events is best fitted with emission measures (EMs) between $4\times10^{45}$ and $2\times10^{47}\, \mathrm{cm^{-3}}$ with temperatures from 10.7 to 15.5\,MK \citep{hannah2008}.
 
 Due to the low-energy limit of the STIX response of around 4\,keV and because the thermal bremsstrahlung emission process is highly temperature dependent, STIX is only sensitive to plasma above $\sim$\,8\,MK. Furthermore, the indirect imaging concept used by STIX is less sensitive than focusing instruments and requires good count statistics to be able to reconstruct X-ray images \citep{stix2020V2}. For very small events, imaging may be limited and one has to rely solely on the spatially integrated X-ray spectrum. These limitations can however partly be mitigated by using complementary data from the narrow SDO/AIA UV/EUV filters to provide high-resolution imagery of the flaring corona and chromosphere.
 
 The microflares studied here occurred in an active region that was also visible to Earth orbiting instruments, providing this opportunity to perform multi instrument analysis. The AIA filters cover a temperature range from 20 000\,K up to 20\,MK supplementing STIX observations of thermal plasma with some overlap \citep{Lemen2012}. EUV imagers are also carried by the Solar Orbiter spacecraft, the Extreme Ultraviolet Imager (EUI) suite consisting of three telescopes, but are only turned on for 30 days for each orbit because of their high telemetry requirements \citep{soloscience2020V2}. While joint observations of the instruments carried by Solar Orbiter promise valuable results because of their vicinity to the Sun, they were not operational during the time of the STIX commissioning flares under study. Even if such data is available, the AIA observations are still valuable due to the multitude of filters which allow for DEM reconstruction not possible with Solar Orbiter imaging data.
 \section{Data and Methods}
 \subsection{STIX}
 \label{dataMethods}
The two flares under study were selected from the set of the 69 microflares that were observed during the STIX commissioning phase between May 18 and June 13, 2020. From this set, 26 were clearly observed in at least two STIX energy channels \citep{battaglia2021}. Based on the background subtracted GOES flux, the events studied here are classified as B2 and B6 events. The flare of GOES class B2 occurred on June 6th, the B6 event on June 7th, 2020. They were selected because the STIX countrates are sufficient to reconstruct X-ray spectra. When STIX observed these events, the Solar Orbiter spacecraft was at a distance of 0.53\,AU from the Sun. The light travel time differences between Solar Orbiter and Earth are 250.3 and 251.8 seconds for the events on June 6th and 7th respectively and are added to the STIX observations in the further analysis.

The spectral fitting was performed using the OSPEX software \citep{schwartz2002} in the SolarSoftware (SSWIDL) package. For the transmission function of the instrument components in the X-ray path, nominal values based on their theoretical properties were assumed. A preliminary version of the STIX energy calibration was used. To isolate the contribution from flaring plasma, the backgrounds from non-flaring times were subtracted. The X-ray spectra were fitted assuming either an isothermal plasma with the ''f\_vth'' function or a combination of isothermal and nonthermal thick target bremsstrahlung emission with the ''f\_thick2'' function in OSPEX.  This fitting was done with an integration time of 12 seconds for the B6 and 20 seconds for the B2 flare to infer the temporal evolution of the nonthermal emission and emission measure as well as temperature for the thermal flare plasma.

Using the total emission measure $\mathrm{EM}$ and temperature $T$ from spectral fitting, the thermal energy was calculated using 
\begin{equation}
\label{Eq-Eth1}
E_{th}=3k_{B}T\sqrt{\mathrm{EM}\,V}
\end{equation}
(e.g. \citealt{aschwanden2015}), with the Boltzmann constant $k_{B}$ and flare volume $V$. The filling factor of the hot emitting plasma is assumed to be 1 \citep{veronig2005}.

Because determination of flare volumes is subject to high uncertainties, we derived a range of plausible volume estimates for both events using different approaches. For an upper limit, we measured the length $L$ and separation $D$ of the flare ribbons from the AIA 171\,\AA\, and 211\,\AA\, images during the peak of the flare (cf. Figs. \ref{f-AIA1}, \ref{f-AIAB2}), and derived the emitting area as $A=L \times D$. From these measurements, the volume was estimated by taking $V=A^{3/2}$. In addition, we estimated the volume of the hot emitting flare plasma from the emitting area estimated by the preliminary STIX amplitude imaging over the energy range 6--10\,keV. As there was no STIX imaging available for the B2 flare on June 6th, the lower boundary was derived assuming that only the shell between the inner and outer boundaries of the flare ribbons connected by semi circles contribute significantly to the emission: $V=\frac{\pi}{2} L (r_{outside}^2-r_{inside}^2)$.  
The ranges of the estimated volumes are $8.6\times10^{26}$ to $8.8\times10^{27}\mathrm{cm^{-3}}$ for the B6 and  $9.4\times10^{25}$ to $8.8\times10^{26}\mathrm{cm^{-3}}$ for the B2 event. 
From the thick target fit, the nonthermal power from the flare electrons was calculated with the \textsf{calc\_nontherm\_electron\_energy\_flux} routine in SSWIDL.
\subsection{AIA}
We used the six coronal EUV filters of AIA (94, 131, 171, 193, 211 and 335\,\AA) for EUV imaging, lightcurves and to reconstruct the differential emission measure (DEM) using the inversion code developed by \cite{HK2012}. These EUV channels image plasma over a temperature range from $10^{5}$\,K to $10^{7}$\,K with a 12\,second cadence and with 0.6\,arcsec pixel resolution \citep{Lemen2012}. In addition, the 1600\,\AA\, filter sensitive to chromospheric $10^{5}$\,K plasma was analysed for the B2 event on June 6th because it shows the response of the chromospheric plasma to the energy deposited there via the thick target deceleration of precipitating flare accelerated electrons. 

The AIA data was processed to Level 1.5 with the \textsf{aia\_prep.pro} SSWIDL routine. Similar to the treatment of the STIX data, pre-event levels were subtracted for the construction of AIA lightcurves.

\subsubsection{Differential Emission Measure Analysis}

When analysing optically thin plasma (as in the solar corona) under the assumption of thermodynamic equilibrium, the differential emission measure along the Line of Sight (LoS) $h$ is commonly defined as
	\begin{equation} 
	\label{Eq-DEM-def}
		\mathrm{DEM(}T\mathrm{)} = n(T)^{2}\frac{\mathrm{d}h}{\mathrm{d}T}.
	\end{equation}
	The $\mathrm{DEM(}T\mathrm{)}$ function is the temperature distribution of the radiating plasma in the LoS and	$n$ denotes the electron number density which is a function of the temperature $T$. The DEM is related to the recorded intensity $I_{\lambda}(T)$ in one of the six AIA filters $\lambda$ via
	\begin{equation}
	\label{Eq-IDEM-def}
		I_{\lambda} =\int_{T} K_{\lambda}(T)\mathrm{DEM(}T\mathrm{)}\mathrm{d}T
	\end{equation}
	with the response function of the filter in question $K_{\lambda}(T)$. This response function depends on the elemental abundances in the emitting plasma, its temperature and the sensitivity of the sensor \citep{HK2012}. The filter response function was calculated from the \textsf{aia\_getresponse.pro} routine in SSWIDL with coronal abundances from the CHIANTI 9 database \citep{chianti97,chianti19}. For better counting statistics, the AIA data was binned 2x2 before the DEM reconstruction, resulting in an effective spatial resolution of 1.2 arcsec. The temperature range for the DEM reconstruction was set from 0.5 to 15\,MK. 
	
	The total emission measure of the observed plasma is calculated by integrating over the temperature range of interest
	\begin{equation}
	\label{Eq-EM-def}
		\mathrm{EM} = \int_{T} \mathrm{DEM(}T\mathrm{)}\mathrm{d}T
	\end{equation}
	which is related to the squared electron density along the LoS. As an estimate of the mean plasma temperature, we calculate the emission weighted temperature as
	\begin{equation}
	\label{Eq-T-def}
		\overline{T} =\frac{\int_{T} \mathrm{DEM(}T\mathrm{)}T\mathrm{d}T}{\mathrm{EM}}
	\end{equation}
\citep{cheng2012,vanninathan2015}. 
To isolate the contribution of the flaring plasma, a background DEM was calculated before the start of the event and subtracted from the flare DEMs individually for each pixel of the DEM maps.

The recovered DEM allows for more accurate estimations of the thermal energy by taking into account the multi-thermal DEM distribution rather than the isothermal assumption used in Eq. \ref{Eq-Eth1}. This is done by summing over the thermal energies in all temperature bins $T_{k}$ of the DEM \citep{aschwanden2015}:
\begin{equation}
\label{Eq-Eth2}
E_{th}=3k_{B}V^{1/2}\displaystyle\sum_{k}^{}T_{k}[\mathrm{DEM}(T_{k}) \Delta T_{k}]^{1/2}. 
\end{equation}
For calculating spatially integrated quantities such as average temperature, the reconstructed DEMs from all pixels inside a box covering the event were summed into a single DEM distribution characterizing the flaring plasma. 
\subsection{GOES}
For better statistics, the data from the 0.5--4 and 1--8\,\AA\, SXR bands from the GOES-16 satellite (available at 1 second cadence) was integrated into 15 seconds time bins. Analogous to the STIX and AIA data processing, the curves were background subtracted by a pre-event level in order to separate the emission from the flaring plasma. From the two SXR bands, EM and temperature were calculated using available SSWIDL routines \citep{goesEMT2005}. 

The thermal energies from GOES data were calculated from Eq. \ref{Eq-Eth1} as the two GOES SXR channels only allow for isothermal EM and temperature reconstructions. The analysis of the Neupert effect involves the calculation of the time derivative of the GOES SXR fluxes. For the small flares under study, the GOES data are noisy and thus the derivatives are very fluctuating. Therefore, for robustness we first smoothed the data using the algorithm described in \cite{Podladchikova2017} before calculating the derivatives.

\section{Results}
\subsection{B6 Flare on 21:42\,UT, June 7, 2020}
\begin{figure}
\includegraphics[width=0.48\textwidth]{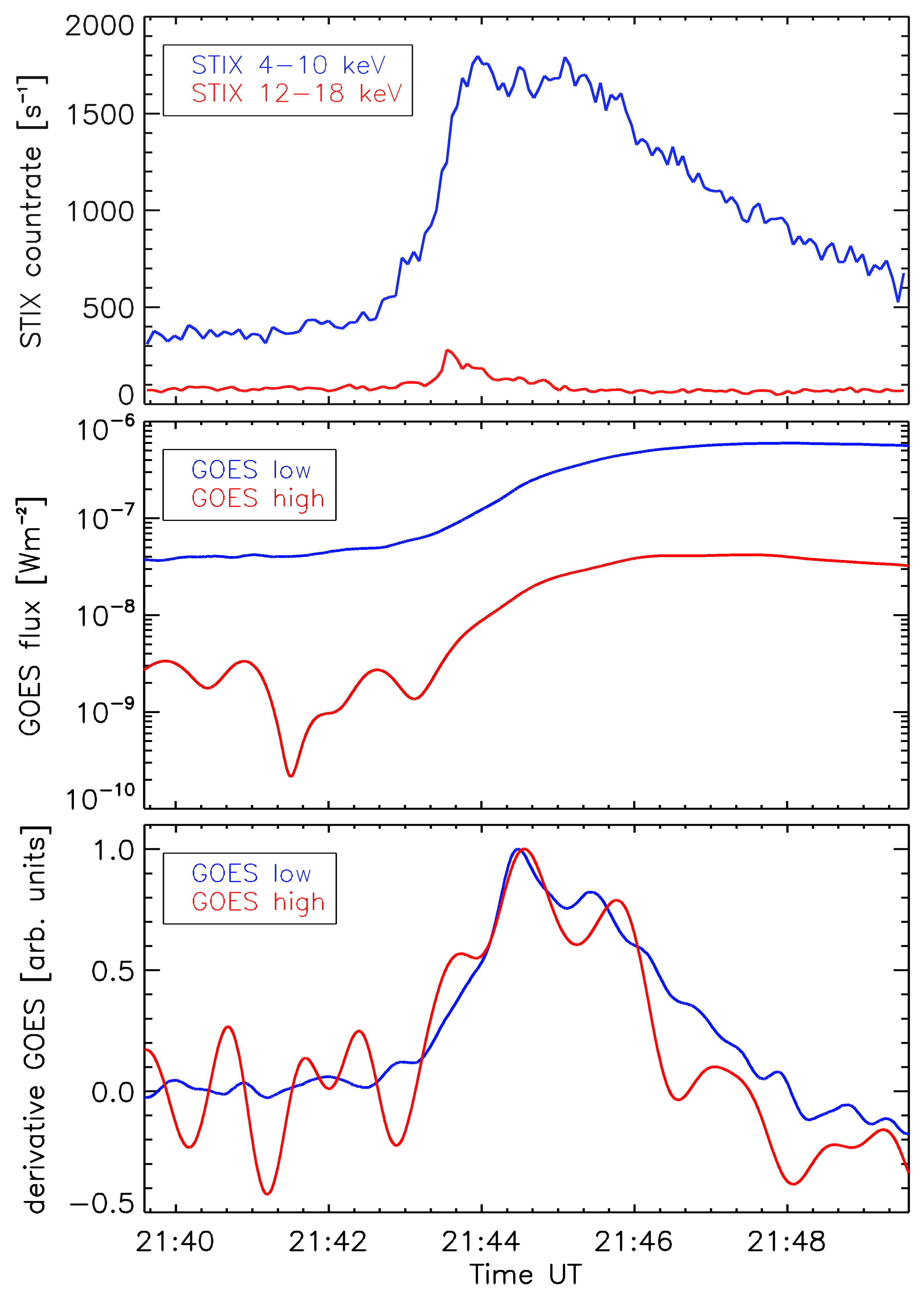}
\caption{Light curves during the June 7, 2020 B6 flare. Top: STIX X-rays at 4--10 and 12--18 keV. Middle: smoothed GOES 0.5--4 (``high'') and 1--8\,\AA\,(``low'') lightcurves. Bottom: Time derivatives of both GOES channels.}
\label{f-LQs1B6}
\end{figure}

Figure \ref{f-LQs1B6} shows X-ray lightcurves in two STIX energy bands, the two GOES SXR bands and the corresponding GOES SXR flux derivatives. The STIX HXRs in the 12--18\,keV band show an impulsive peak centred at 21:43:35 UT while the lower energy 4--10 keV band shows a more gradual time profile as it is sensitive to the thermal flare plasma. The gap between the two energy bands is intended to provide a cleaner break between thermal and nonthermal components. The peak in the STIX 12--18\,keV energy bin roughly coincides with the first peak in the derivative of the GOES 0.5--4\,\AA\, channel while the main peaks in the derivatives of both GOES channels occur one minute after the maximum in the STIX HXR data.
\begin{figure}
\includegraphics[width=0.45\textwidth]{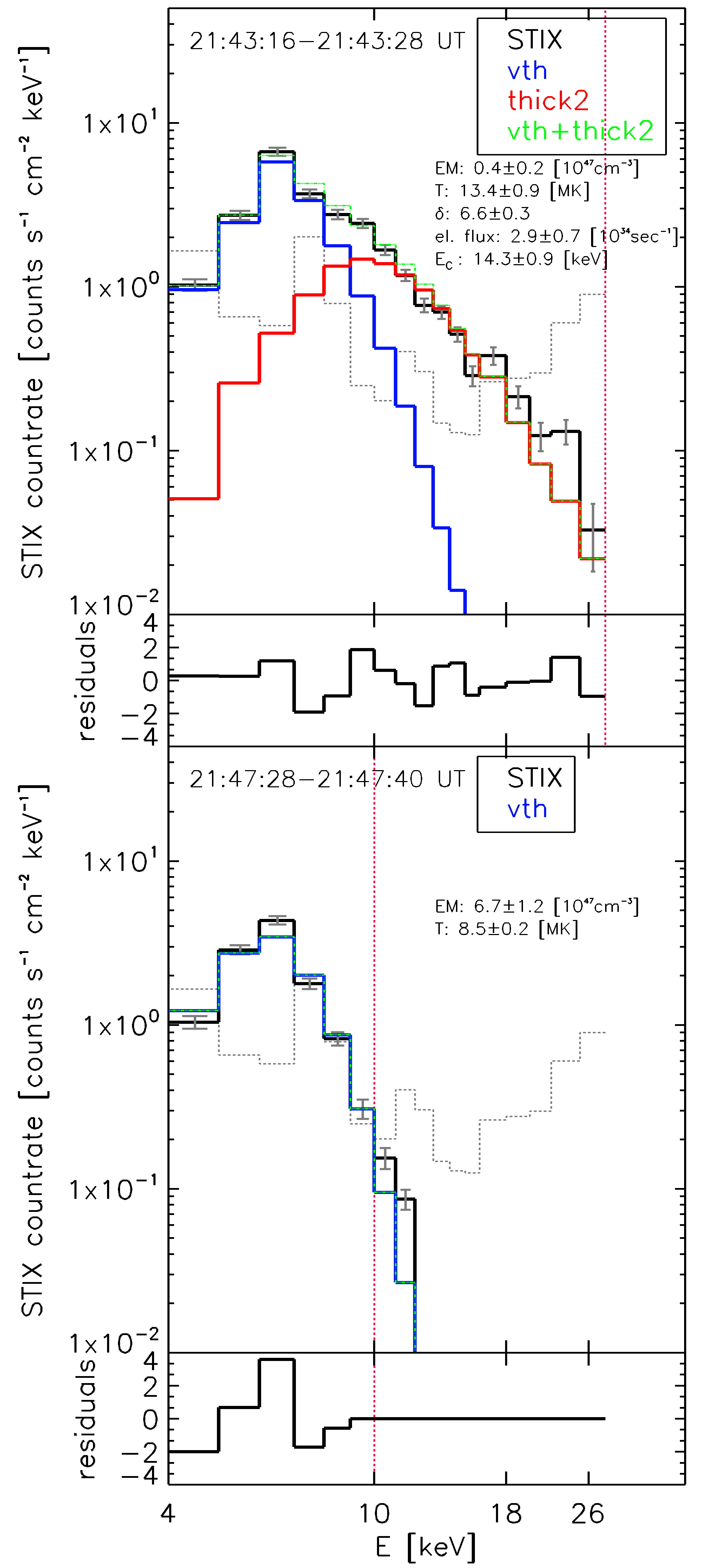}
\caption{STIX X-ray spectra for the B6 event on June 7th, 2020, during the impulsive phase (top panel) and gradual phase (bottom panel). Black: background subtracted data. Blue: isothermal fit.  Red: thick target fit. Green: sum of isothermal and thick target emission. Grey: Preflare background. Red vertical lines indicate the energy range used for fitting (4--28 and 4--10 \,keV respectively).}
\label{f-spectraB6}
\end{figure}

Figure \ref{f-spectraB6} shows STIX X-ray spectra at two distinct time steps during the impulsive and gradual flare phase along with the best fits assuming thermal and thick target bremsstrahlung emissions. During the flare impulsive phase from 21:42\,UT to 21:45\,UT, all spectra are best fit assuming a combination of a thermal and a nonthermal thick target single power-law. The top panel of Fig. \ref{f-spectraB6} shows a 12-second integrated spectrum from this time period. The fitted thick target emission has a steep power law index of $\mathrm{\delta=}$\,($6.6 \pm 0.3 $) above a low energy cutoff $\mathrm{E_{C}=}$ ($14.3\pm0.9$) keV and an electron flux of ($2.9 \pm 0.7$)\,$\mathrm{10^{34}\,sec^{-1}}$. During this impulsive phase, the contribution from thermal plasma using an isothermal fit is quantified to be $\mathrm{EM}=$\,($0.4\pm 0.2$)\,$\mathrm{10^{47}\,cm^{-3}}$ with an isothermal temperature of $\mathrm{T}=$\,($13.4\pm0.9$)\,$\mathrm{MK}$. During the gradual phase (bottom panel of Fig. \ref{f-spectraB6}), the best fit is achieved without any nonthermal emission and an isothermal component with $\mathrm{EM}=$\,($6.7\pm 1.2$)\,$\mathrm{10^{47}\,cm^{-3}}$ and $\mathrm{T}=$\,($8.5\pm0.2$)\,$\mathrm{MK}$. This represents an order of magnitude higher emission measure compared to the impulsive phase.

\begin{figure*}
\includegraphics[width=0.98\textwidth]{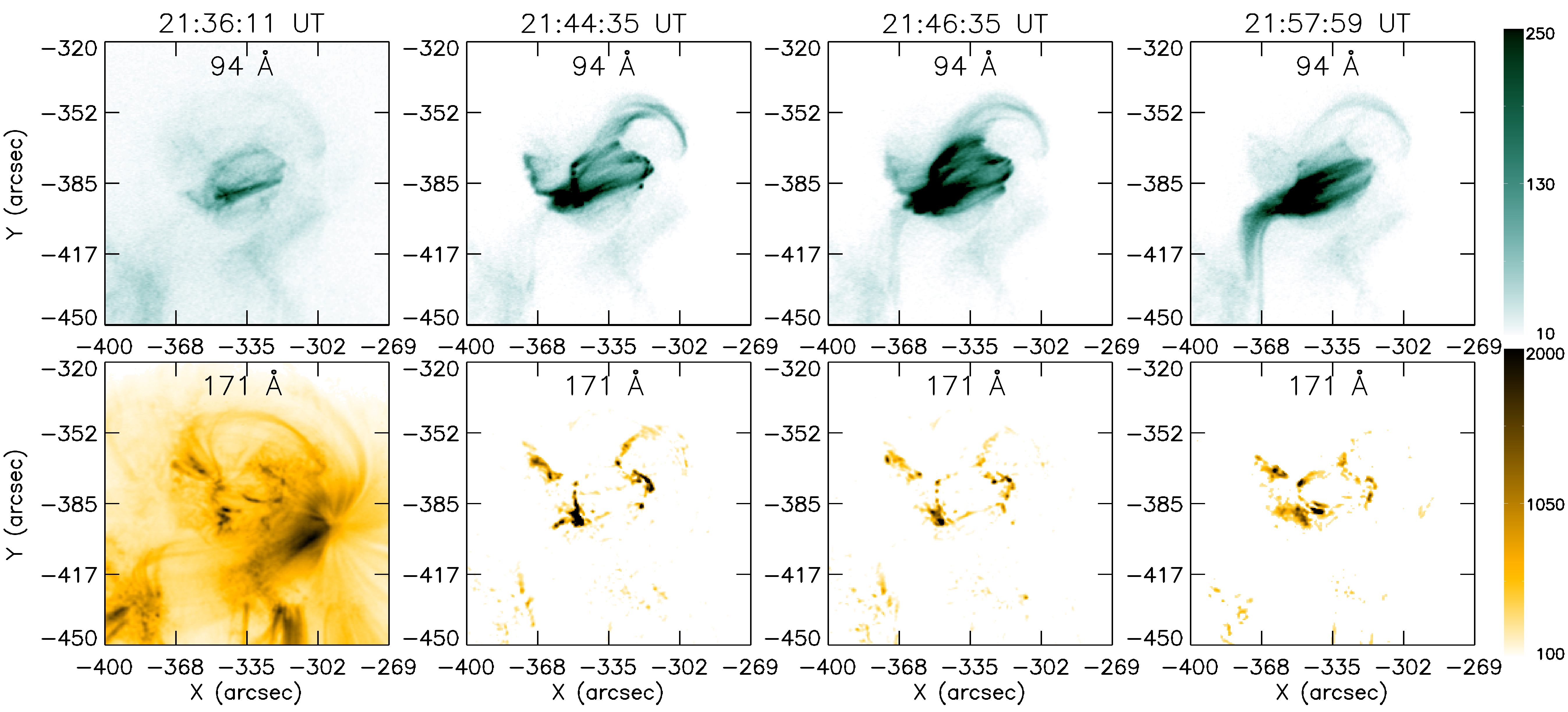}
\caption{AIA 94 and 171\AA\,
images for the B6 flare on June 7th, 2020. From left to right: Preflare configuration, impulsive and gradual phase. The flaring 171\AA\, images are background subtracted. Colour bars indicate the scaling used in the three rightmost panels.}
\label{f-AIA1}
\end{figure*}

Figure \ref{f-AIA1} shows AIA EUV images in the 94 and 171\,\AA\, filters. For clarity, the background has been subtracted in the three 171\,\AA\, panels.
Compared with the pre-event image shown in the leftmost panel, during the flare evolution the 94\,\AA\, images (Fe XXI, with a peak formation temperature $T\sim7$\,MK) show the filling of the loops by heated plasma at 21:44:35\,UT and a further increase at 21:46:35\,UT. The snapshot at 21:57:59\,UT shows the decay phase of the flare, where the hot emission sampled in the 94\,\AA\, filter is decreasing again. In the 171\,\AA\, images the flare kernels can be identified and are most pronounced at 21:44\,UT, which coincides with the peak in the STIX high-energy bands (see 12--18 keV light curve in Fig. \ref{f-LQs1B6}) and the nonthermal component identified in the STIX spectrum at that time (Fig. \ref{f-spectraB6}, upper panel).
\par
\begin{figure*}
\includegraphics[width=0.98\textwidth]{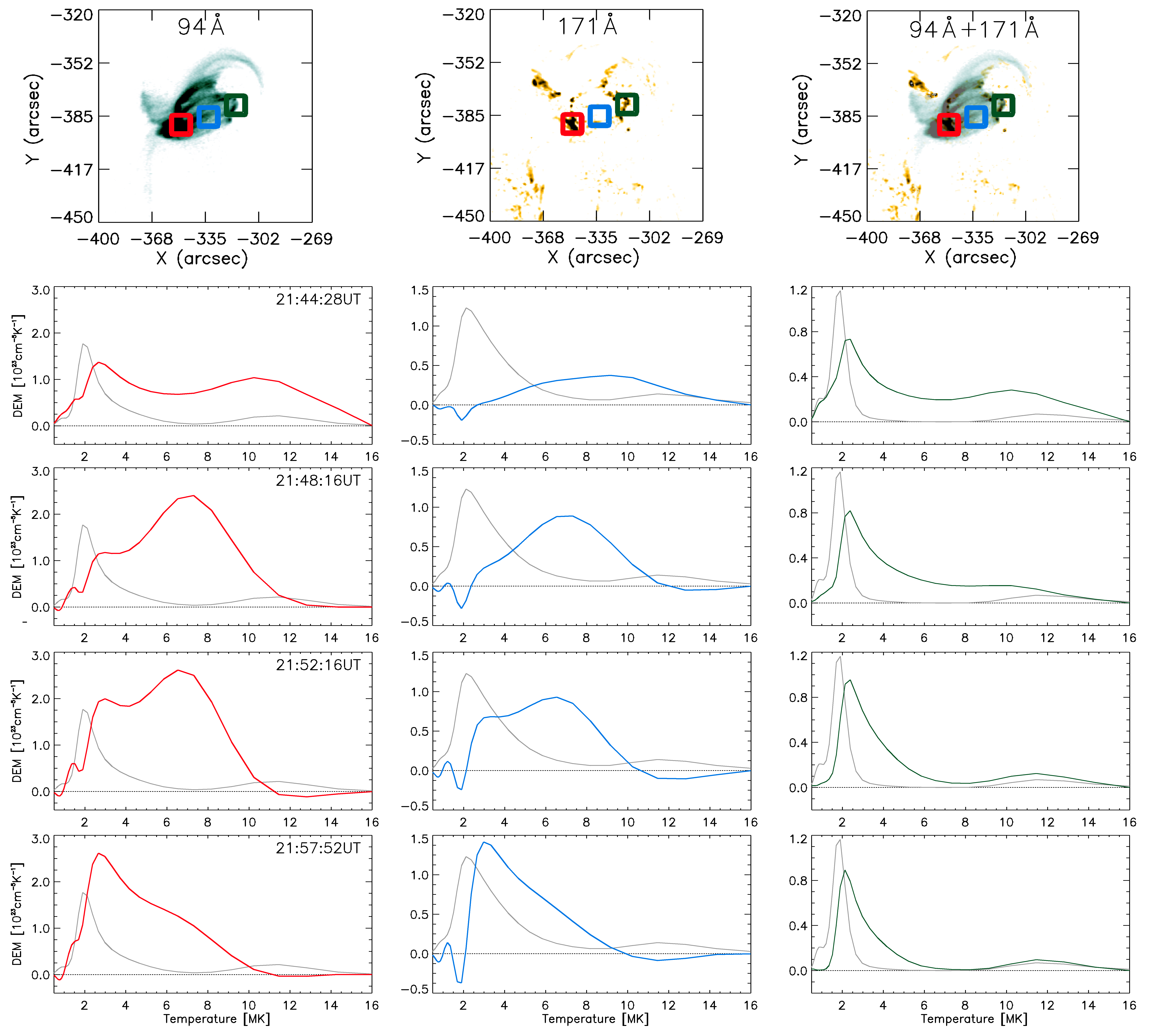}
\caption{Top: AIA 94 and 171\,\AA\, images as well as a composite of both channels taken during the B6 flare on June 7th, 2020 around 21:44\,UT. Coloured boxes indicate the regions used for extracting the DEM curves shown in the bottom panels. 
Bottom: Time evolution of the preflare subtracted DEM profiles averaged over the boxes shown in the maps in the top panels. Black curves show the corresponding preflare DEMs.}
\label{f-DEMProfilesB6}
\end{figure*}

The distinction between flare loops and kernels is further studied by analysing the spatio-temporal evolution of the derived DEM distributions and maps. Figure \ref{f-DEMProfilesB6} shows snapshots of the time evolution of the reconstructed DEM profiles from different regions.
The time evolution of the DEM profile calculated inside a box covering flare loops is shown in the middle panels in blue. The pre-flare background DEM distribution in grey peaks at roughly 2\,MK and extends to over 4\,MK. After the flare onset at 21:44\,UT, a broad hot component around 10--12\,MK starts to appear. The loops gradually fill with hot plasma with the DEM profile centred roughly at 7\,MK at 21:48\,UT. At 21:52\,UT, the loop plasma cools down and a cooler component centred around 3\,MK develops. By 21:57\,UT, most of the plasma in the flare loops has cooled down to around 3\,MK as seen in the bottom panel. 

The pre-flare DEM profile of the western flare kernel shown in green in the right panels  of Fig. \ref{f-DEMProfilesB6} peaks at roughly 2\,MK similar to the flare loop background but is much narrower with virtually no plasma above 3\,MK. After the flare onset, the western flare kernel shows a $\sim 3$\,MK component developing that does not cool significantly during the 14 minute evolution shown in Fig. \ref{f-DEMProfilesB6}. The eastern kernel (left panels) shows a similar albeit broader pre-event background DEM with a peak around 2\,MK. Over the course of the event, a $\sim 3$\,MK component develops similarly to the western kernel. In addition, a hot 7\,MK component that subsequently cools down is observed. This behaviour of the eastern kernel that shows a combination of the western kernel and the isolated flare loops is probably due to loops as well as footpoints being present along the line of sight. 

\begin{figure}
\includegraphics[width=0.48\textwidth]{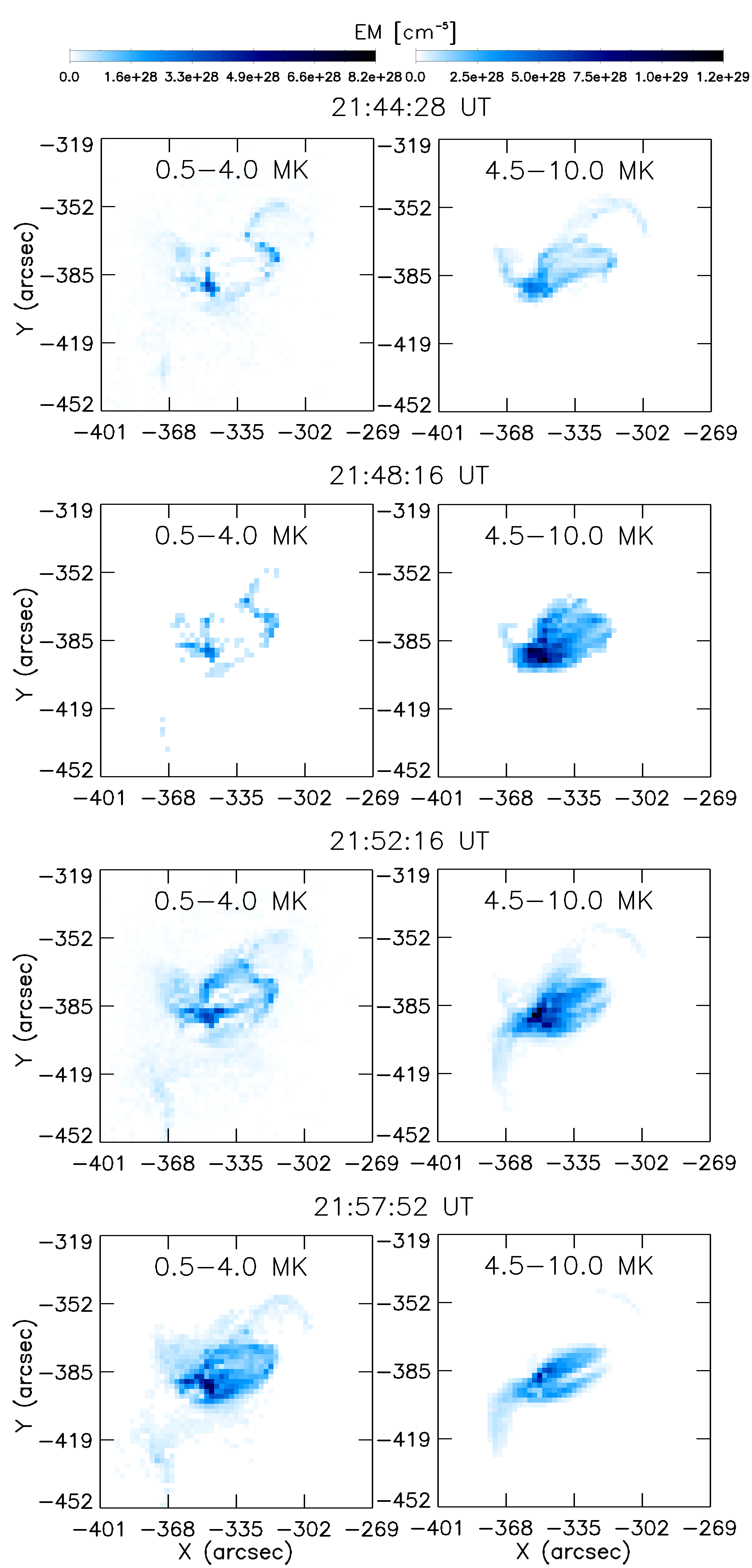}
\caption{EM maps in temperature bins from 0.5 to 4.0 and 4.5 to 10\,MK reconstructed from the AIA DEM over the course of the B6 flare on June 7th, 2020, highlighting the cooler flare kernels and hotter loops.}
\label{f-EMBinsB6}
\end{figure}

This different response and temperature of the plasma in the flare loops and flare kernels is further illustrated in Fig. \ref{f-EMBinsB6}.
Here maps of the total EM integrated from 0.5 to 4\,MK (left column) and from 4 to 10\,MK (right column) are shown for the same times as the DEM curves in Fig. \ref{f-DEMProfilesB6}. At 21:44\,UT and at 21:48\,UT, the cooler EM bin clearly shows the flare kernels while the hotter bin shows the hot plasma in the loops. At 21:52\,UT and more pronounced at 21:57\,UT the flare loops can now also be seen in the 0.5--4\,MK bin while there is less EM in the hot 4.5--10\,MK bin compared to 21:48\,UT as the thermal plasma in the loops cools down. 
\begin{figure}
\includegraphics[width=0.48\textwidth]{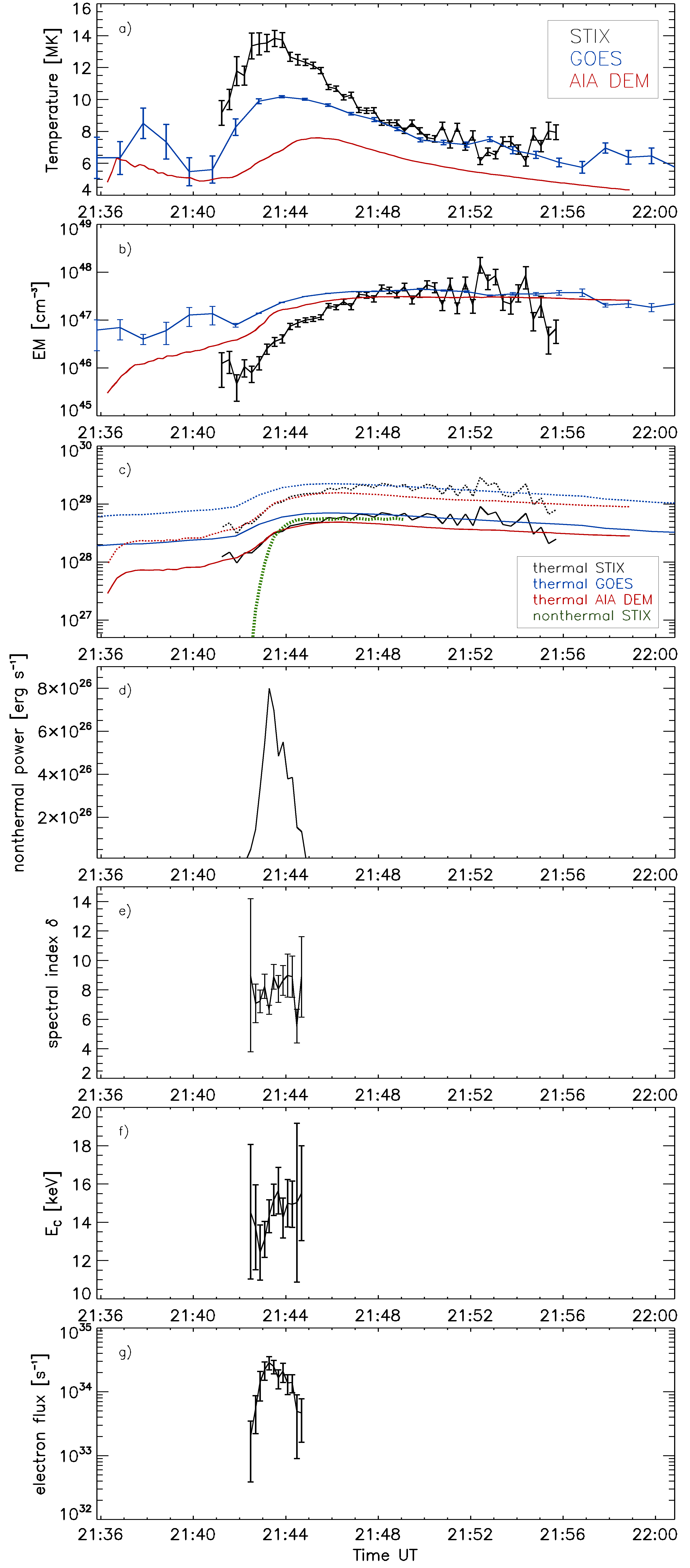}
\caption{Top panels: Time evolution of temperature, EM and thermal energy of the flaring plasma derived from STIX, GOES and AIA data for the B6 flare on June 7th, 2020. For the thermal energies, upper (dashed lines) and lower (solid lines) estimates using different flare volumes are shown. Bottom panels: Evolution of the nonthermal energy and parameters from the thick target fit to the STIX spectra during the flare impulsive phase.}
\label{f-fitParB6}
\end{figure}

The spatially integrated results from the DEM reconstructed using AIA EUV data are combined with STIX and GOES observations in Fig. \ref{f-fitParB6}. The panels a) and b) show the plasma temperature and EM derived from STIX, GOES and the reconstructed AIA DEM. The three methods yield different results in terms of absolute values as well as timing. The flare plasma temperature peaks first in the STIX data, while the temperature derived from the AIA DEM lags behind about 3\,minutes. Peak plasma temperatures lie between 7\,MK (AIA) and 14\,MK (STIX). The flare as observed with STIX is preceded by a rise in the plasma temperature seen by GOES (around 21:38 UT). EM values range from $\mathrm{10^{46}}$ to $\mathrm{10^{48}\,cm^{-3}}$. Panel c) shows the thermal energies calculated from the different instruments (using Eq. \ref{Eq-Eth1} for STIX/GOES and Eq. \ref{Eq-Eth2} for AIA) using the upper (dashed lines) and lower (solid lines) volume estimates as discussed in Section \ref{dataMethods}. The green line shows the time integral of the nonthermal energy rate derived from the thick target fit shown in panel d). The cumulated energy in nonthermal electrons during the flare impulsive phase roughly accounts for the thermal energies derived from STIX, GOES and AIA when using the lower bound of the volume estimate (derived from STIX imaging), but it is up to an order of magnitude smaller when compared to the upper limits. Panels d) to g) show the power of the electrons and the fit parameters for the STIX nonthermal component during the impulsive flare phase between 21:42\,UT and 21:45\,UT. Over the course of the impulsive phase, the injected nonthermal electron flux is characterized by a rather steep power-law index that fluctuates around $\delta=8$ above a low energy cutoff $\mathrm{E_{C}}$ that varies between 12 to 16 \,keV. The deduced electron flux is of the order of a few $10^{34}\mathrm{s^{-1}}$.
\subsection{B2 Flare on 19:14\,UT, June 6, 2020}
\begin{figure}
\includegraphics[width=0.48\textwidth]{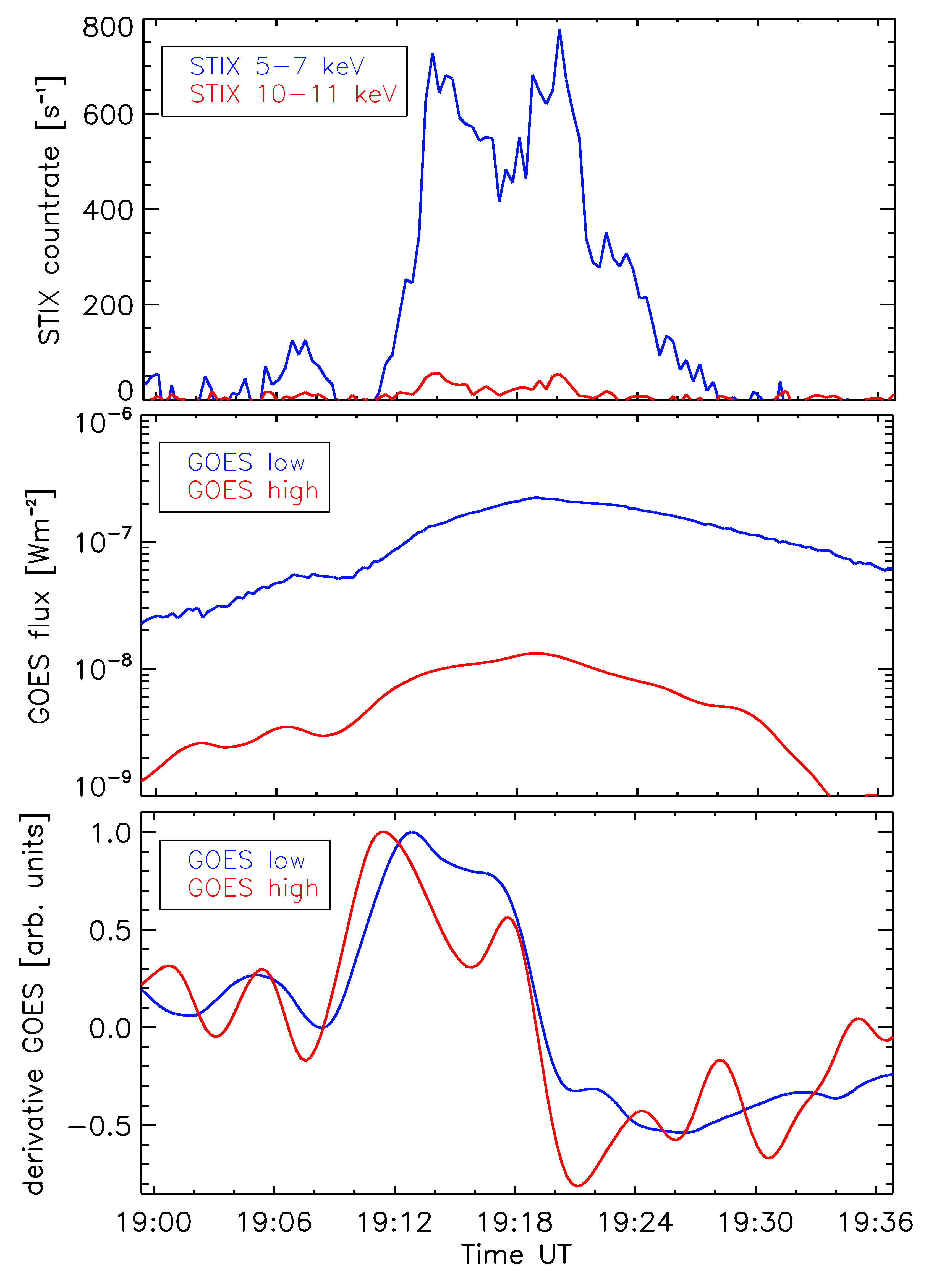}
\caption{Top: Background subtracted lightcurves in different STIX energy bands for the B2 event on June 6th, 2020. Middle: smoothed GOES 0.5--4 (``high'') and 1--8\,\AA\,(``low'') lightcurves. Bottom: Derivatives of both GOES channels.}
\label{f-LQs1B2}
\end{figure}

\begin{figure}
\includegraphics[width=0.48\textwidth]{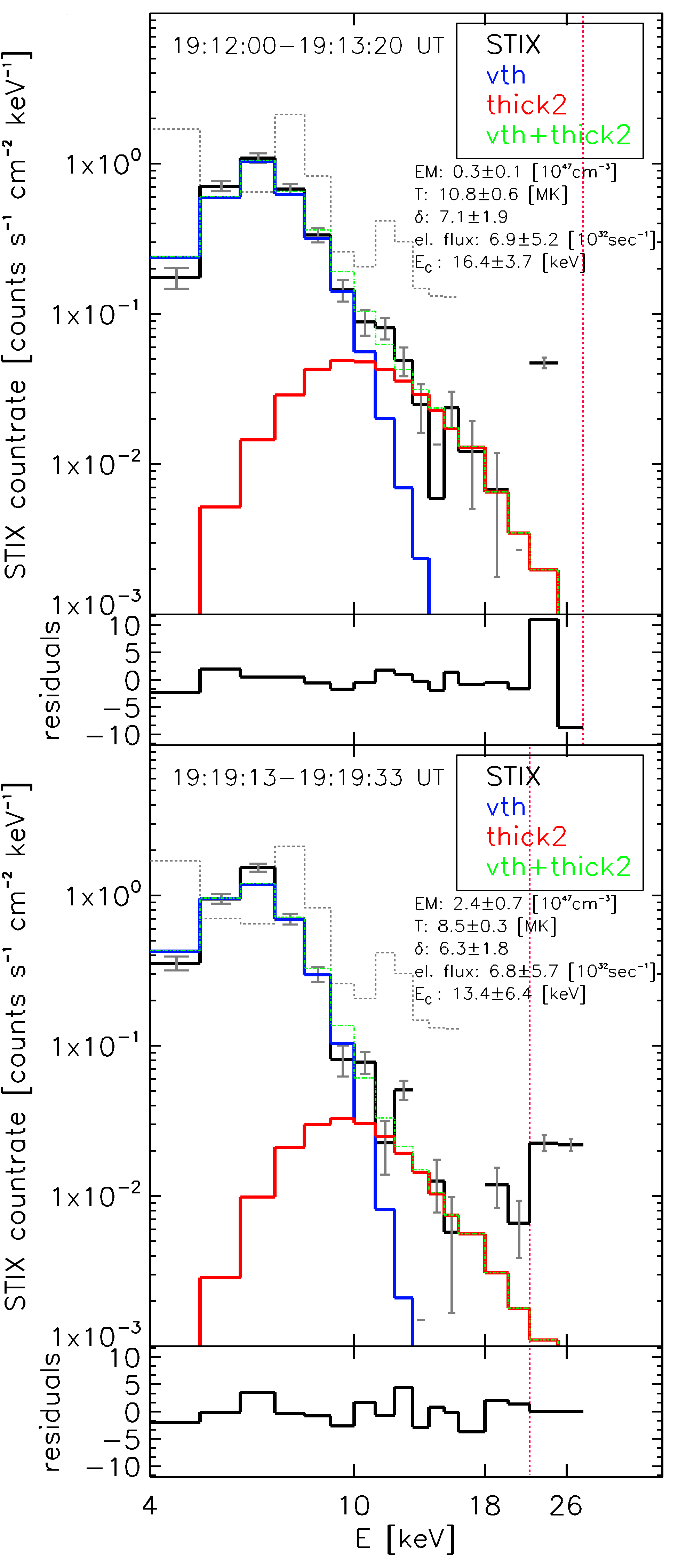}
\caption{STIX X-ray spectra for the B2 event on June 6th, 2020 during the impulsive phase (top panel) and gradual phase (bottom panel). Black: background subtracted data. Blue: isothermal fit.  Red: thick target fit. Green: sum of isothermal and thick target emission. Grey: Preflare background. Red vertical lines indicate the energy range used for fitting (4--28 and 4--22 \,keV respectively).}
\label{f-spectraB2}
\end{figure}

The top panel of Fig.\ref{f-LQs1B2} shows STIX lightcurves for the 5--7, and 10--11 keV channels. They exhibit a pronounced double-peak structure. The middle panel shows the GOES SXR fluxes and the bottom panel the derivatives of the data. The GOES derivatives show a main peak in association with the first STIX burst, the 0.5--4\,\AA\, channel shows also a second (smaller) rise along with the second STIX peak. Although the timing of the HXR peaks and SXR derivatives are not perfect, they indicate a basic Neupert effect agreement for both peaks.

Figure \ref{f-spectraB2} shows fits to STIX spectra about 6 minutes apart that roughly coincide with the two peaks in the STIX lightcurves at around 19:13\,UT and 19:19\,UT. Both spectra show indications of nonthermal emission, and are therefore fitted with a thermal + thick-target two-component fit. Comparison to the determined background level in grey shows that the flare enhancement above the background is smaller than the background level itself. The background however is stable and predictable over longer timescales because it is dominated by the radioactive calibration sources of the instrument \citep{battaglia2021}. At 19:13\,UT, the spectral fitting yields thermal plasma with an EM $=$($0.3\pm 0.1)$\,$\mathrm{10^{47}\,cm^{-3}}$ and a temperature T $=$($10.8\pm0.6$)\,$\mathrm{MK}$. The thick target emission is characterized by a power law index of $\mathrm{\delta}=$\,($7.1 \pm 1.9 $) above a low energy cutoff $\mathrm{E_{C}=}$ ($16.4\pm3.7$) and an electron flux of ($6.9 \pm 5.2$)\,$\mathrm{10^{32}\,sec^{-1}}$. 6 minutes later at 19:19\,UT, the spectrum shows significantly more thermal plasma with an EM $=$($2.4\pm 0.7$)\,$\mathrm{10^{47}\,cm^{-3}}$ and T$=$($8.5\pm0.3$)\,$\mathrm{MK}$. The power law index is $\mathrm{\delta}=$\,($6.3 \pm 1.8$) above a low energy cutoff $\mathrm{E_{C}=}$ ($13.4\pm6.4$) and the electron flux is similar as 6 minutes before with ($6.8 \pm 5.7$)\,$ \mathrm{10^{32}\,sec^{-1}}$.

\begin{figure*}
\includegraphics[width=0.98\textwidth]{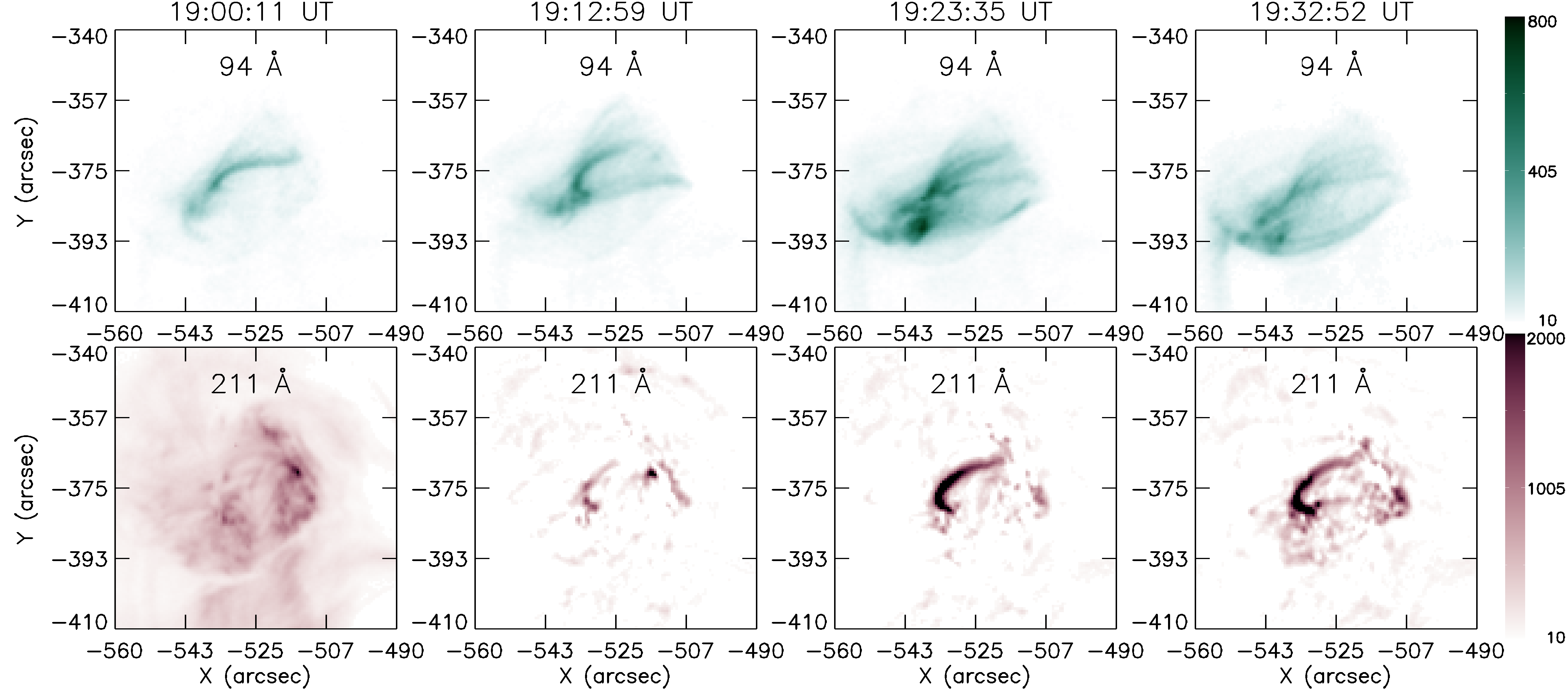}
\caption{AIA 94 on the top and 211\AA\,images on the bottom for the B2 flare on June 6th, 2020. From left to right: Preflare configuration, impulsive and gradual phase. The flaring 211\AA\, images are background subtracted. Colour bars indicate the scaling used in the three rightmost panels.}
\label{f-AIAB2}
\end{figure*}

Figure \ref{f-AIAB2} shows a sequence of AIA 94\,\AA\, and 211\,\AA\,images over the course of the event. In the AIA 94\,\AA\, images, there is already a hot loop present at 19:00\,UT before the onset of the event as observed by STIX and GOES. 
As this initial loop cools down, it subsequently appears in the AIA 211\,\AA\, filter at around 19:13\,UT. Coinciding with the first peak observed in the STIX count rates (Fig. \ref{f-LQs1B2}), more loop plasma starts to appear in the hot 94\,\AA\, channel from 19:13\,UT onwards. Simultaneously, in the 211\,\AA\,filter the flare kernels appear. 

\begin{figure}
\includegraphics[width=0.48\textwidth]{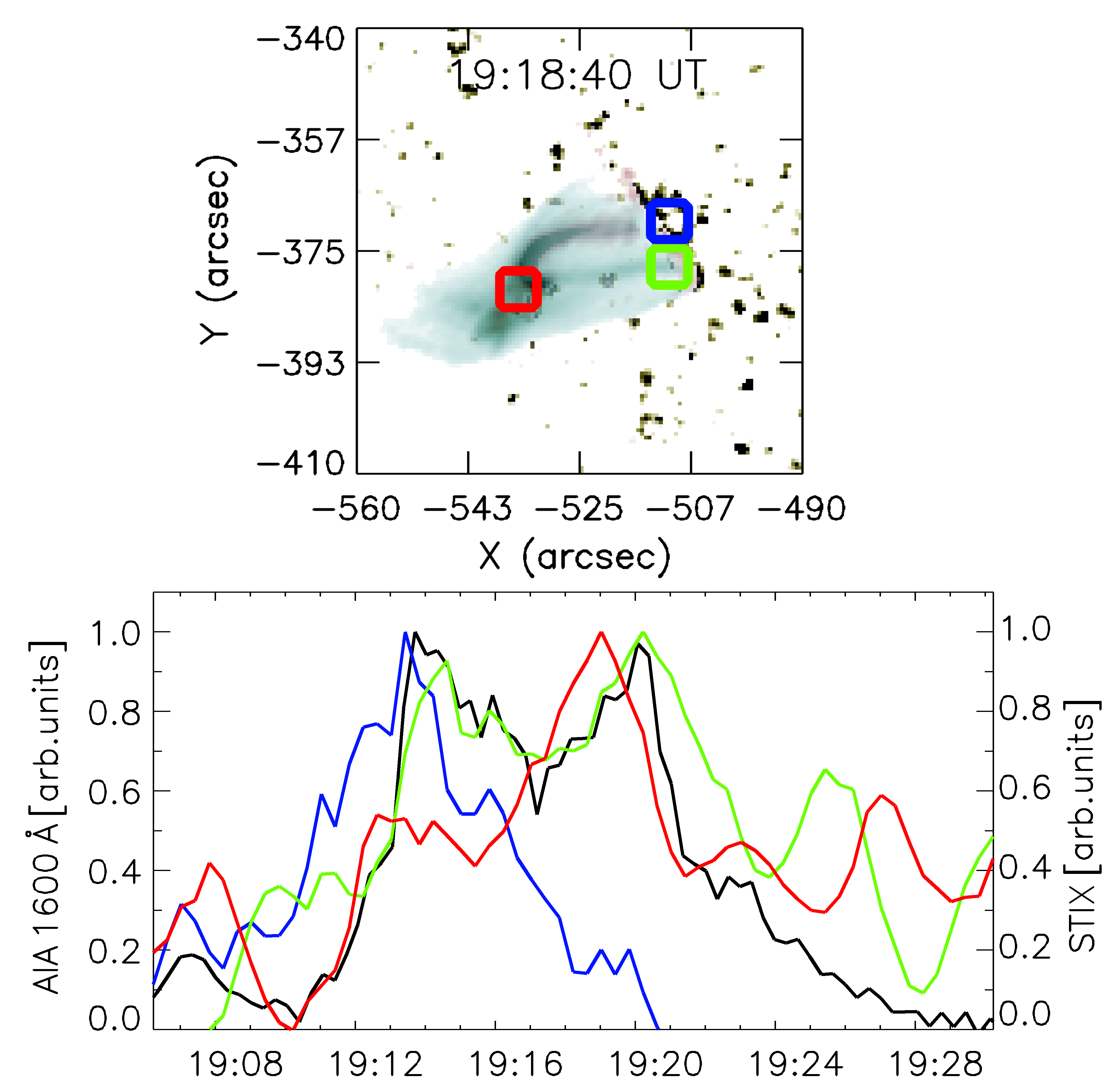}
\caption{Top: Composite image of AIA 94, 211 and 1600\,\AA\,images. The coloured boxes show the areas used for the lightcurves and DEM profiles shown in Fig. \ref{f-DEMProfilesB2}. Bottom panel: AIA 1600\,\AA\,lightcurves from the indicated regions together with the STIX 4-11\,keV light curve (black).}
\label{f-1600LqsB2}
\end{figure}

Figure \ref{f-1600LqsB2} shows a composite image of the AIA 94, 211 and 1600\,\AA\,channels together with 1600\,\AA\, lightcurves extracted from the areas indicated by the coloured boxes.
The chromospheric response in AIA 1600\,\AA\,from the region indicated in yellow, exhibits a double peak shape that closely resembles the STIX profile summed over all energies shown in black. The curves from the blue and red boxes each correspond to one of the two STIX peaks. This indicates that different flaring domains are involved during the two peaks. These two distinct chromospheric responses support the detection of the nonthermal HXR emission below background level from the STIX spectra in Fig. \ref{f-spectraB2}.

\begin{figure*}
\includegraphics[width=0.98\textwidth]{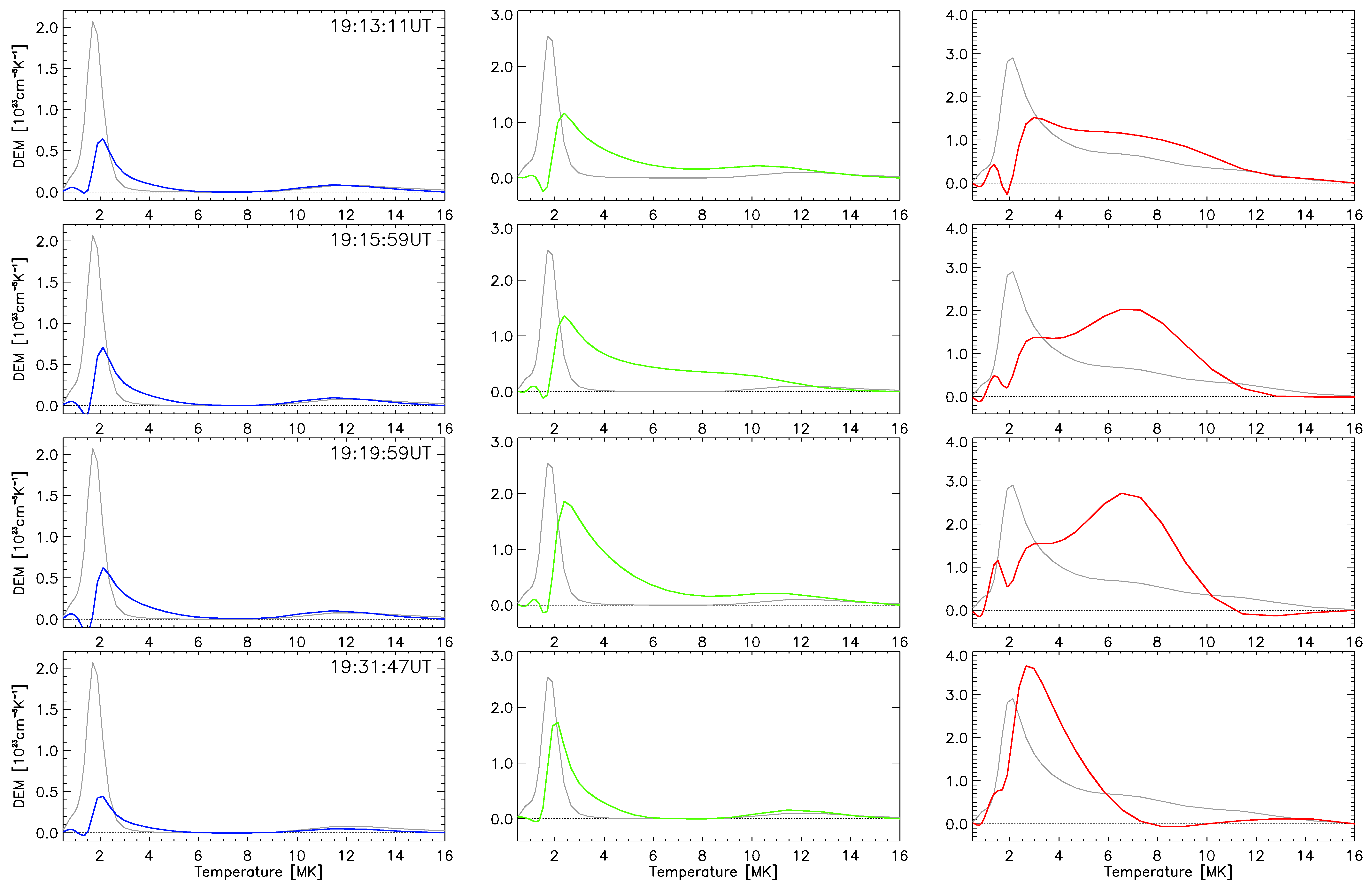}
\caption{Time evolution of the preflare subtracted DEM profiles for the B2 flare on June 6th, 2020 averaged over the boxes shown in the top panel of Fig.\ref{f-1600LqsB2}. Black curves show the corresponding preflare background DEM profiles.}
\label{f-DEMProfilesB2}
\end{figure*}

Figure \ref{f-DEMProfilesB2} shows the evolution of the DEM profiles averaged over the boxes shown in Fig.\ref{f-1600LqsB2} together with the corresponding pre-event DEMs. The pre-event DEM distributions summed over the two western flare kernels shown in the left and middle panels are narrow and centred around 2\,MK. During the flare, DEM distributions centred around 2.5\,MK that gradually subside after the second peak in the STIX X-ray lightcurve (19:20 vs 19:31\,UT) are observed for both western flare kernels.

The DEM curve summed over the eastern kernel shows the same 2.5\,MK component as the western kernels after the event onset. In addition, a broad 7\,MK\,contribution is already present at the start (19:13\,UT) and further increases (19:15\,UT, 19:19\,UT) before cooling down to around 3\,MK (19:31\,UT). As can be seen in the composite image in Fig.\ref{f-1600LqsB2}, flare loop plasma as well as a kernel lie in the LoS for the area marked in red, resulting in the two components of the DEM profile.

\begin{figure}
\includegraphics[width=0.48\textwidth]{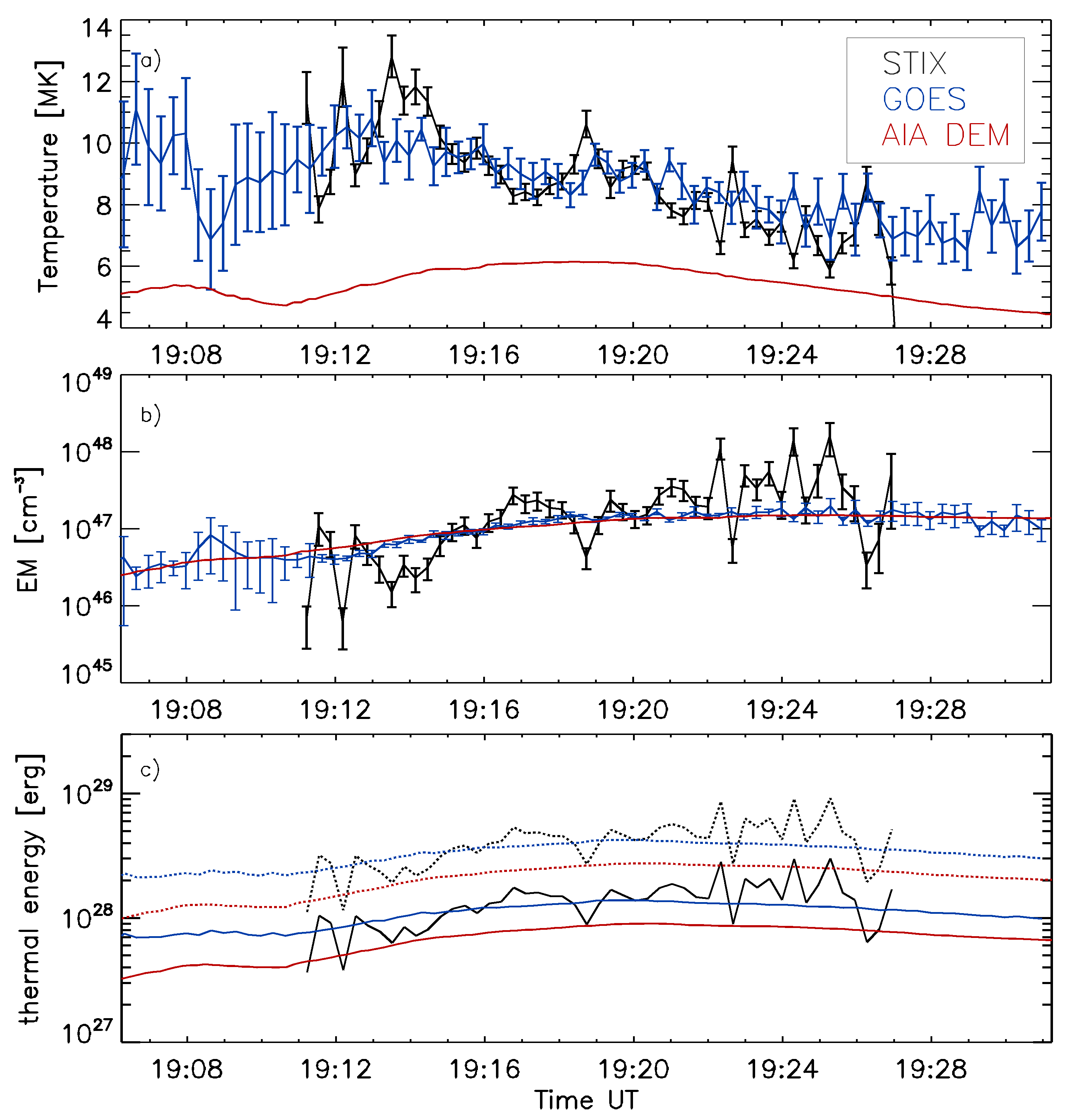}
\caption{From top to bottom: Time evolution of temperature, EM and thermal energy of the flaring plasma derived from STIX, GOES and AIA data for the B2 flare on June 6th, 2020. For the thermal energies, upper (dashed lines) and lower (solid lines) estimates using different flare volumes are shown.}
\label{f-fitParB2}
\end{figure}

In Fig. \ref{f-fitParB2}, the spatially integrated results from the DEM reconstructed from AIA EUV data are compared with STIX and GOES plasma parameters.
The two top panels show the temperature and EM derived from STIX, GOES and the AIA DEM. As in the case of the previously discussed B6 event, the absolute values and timing differ with a time lag between maximum plasma temperature between STIX/GOES and AIA. Values for the EM range from $\mathrm{10^{46}}$ to $\mathrm{10^{48}\,cm^{-3}}$. Peak plasma temperatures lie between 6\,MK (AIA) and 12\,MK (STIX). The thermal energies (bottom panel) derived from STIX and GOES, roughly agree, while for the AIA DEM, they are somewhat smaller. This discrepancy may be related to the AIA EUV filters being sensitive also to cooler plasma along the LoS than STIX and GOES, resulting in a lower mean temperature as shown in the top panel.
\section{Discussion and Conclusions}
We performed case studies for two microflares of GOES class B6 and B2 that were observed on June 7th and 6th 2020 during the STIX commissioning phase. From STIX spectral fitting for the B6 event we find peak plasma temperatures of $\mathrm{T}=$\,14\,MK and emission measures up to $\mathrm{EM}=\mathrm{10^{48}\,cm^{-3}}$. The power-law index of the thick target fit is around $\delta=8$ above low energy cutoffs $\mathrm{E_{C}}$ between 12 to 16 \,keV. The nonthermal electron flux is a few $10^{34}\mathrm{s^{-1}}$. Since $\mathrm{E_{C}}$  derived from the fitting gives an upper limit of the low energy cutoff, the resulting fit parameter therefore provide a lower limit of the power in nonthermal electrons.
For the B2 flare on June 6th, the plasma seen by STIX is characterized by a peak plasma temperature of around $\mathrm{T}=$\,12\,MK and a maximum emission measure of $\mathrm{EM}=\mathrm{10^{48}\,cm^{-3}}$. For the two instances where a nonthermal component could be fitted to the STIX spectra for this event, the power-law indices are $\delta=7.1$ and $\delta=6.3$ above cutoff energies $\mathrm{E_{C}}$ of 16 and 13\,keV. Electron fluxes are around $7\times10^{32}\mathrm{s^{-1}}$ in both instances. These values for the power law index and plasma temperature for both events agree with the typical parameter ranges for microflares of $\delta=$\,4--10 and $\mathrm{T}=$\,10.7--15.5\,MK given by \cite{hannah2008} based on over 9100 RHESSI microflares. The maximum emission measures for both events which reach $\mathrm{EM}=\mathrm{10^{48}\,cm^{-3}}$ are higher than the $\mathrm{EM}=2\times\mathrm{10^{47}\,cm^{-3}}$ reported in \cite{hannah2008} but uncertainties are high. The peak thermal energies using the lower volume estimates are a few $\mathrm{10^{28}}$\,erg for both events which is also consistent with previous RHESSI microflares (\citealt{stoiser2007}, \citealt{christe2014}, \citealt{warmuth2020}).

Both events under study roughly follow the Neupert effect and reveal nonthermal emission in the STIX spectra during the impulsive phase, indicative of electron beams accelerated during the two events. For the B6 event on June 7th, we find that the energy from the nonthermal electrons can roughly account for the thermal energy requirement using the lower volume estimate. For the B2 flare, such a comparison was not possible due to the low flare signal.

 Comparing the thermal plasma properties deduced from STIX, GOES and the AIA DEM we find that the absolute values and timing differ with the highest plasma temperature derived from STIX, followed by GOES and AIA. Since STIX is only sensitive to thermal plasma roughly above 8\,MK while the AIA DEM images plasma down to 0.5\,MK, such differences are not surprising. The temperature response of the GOES SXR instrument lies in between those two extremes, resulting in derived temperatures for the thermal flare plasma that lie between these two extremes. We find that increases in the plasma temperature calculated from AIA lag behind STIX and GOES. We interpret this to occur due to the broad AIA temperature response, the heated flare plasma is initially outside of AIA's temperature sensitivity and subsequently appears in the AIA filters when it cools down slightly.

 While the temperatures derived from the different instruments show substantial differences, the thermal energies are similar. It is noted that the average plasma temperatures inferred from the AIA DEM shown in Figs. \ref{f-fitParB6} and \ref{f-fitParB2} must be interpreted with caution. Due to the broad AIA DEM distributions that include cool (down to 0.5\,MK) as well as hot plasma (> 10\,MK),\,\,the average temperatures derived are low although there is also a significant contribution from hotter flare plasma. Because the calculation of the thermal energy content (Eq. \ref{Eq-Eth2}) takes this multi-thermal distribution into account, the thermal energies from the AIA DEM and X-ray observations agree more closely than naively comparing the total EMs and temperatures would suggest.
 
 To fully characterize the DEM of the flaring plasma over its entire temperature range, it will be awarding for future studies to combine the AIA EUV data which covers the lower temperature range down to around 0.5\,MK with STIX X-ray data which provides information about the hottest thermal plasma. Similar studies of deriving DEMs by combining different instruments have been performed in the past, by for example combining AIA with RHESSI (e.g. \citealt{inglis2014,battaglia2015}), EVE with RHESSI (e.g. \citealt{caspi2014,mctiernan2019}), or AIA, Hinode/XRT and NuSTAR \citep{wright2017}.

The spatio-temporal behaviour of the AIA DEMs is similar for both events. In both flares, the emission measure and temperature in the loops gradually increases with a broad multi-thermal DEM distribution between 6 and 10\,MK, whereas the footpoints show a more impulsive response in the DEM evolution. 
This spatio-temporal behaviour deduced from  the AIA DEM maps coincides with the plasma evolution for the chromospheric evaporation scenario, and is also corroborated by the observed Neupert effect. This finding from the AIA DEM analysis is important for the understanding of the energy transport in microflares, in particular as so far there is only very limited spectroscopic observations of chromospheric evaporation flows for microflares available (\citealt{stoiser2009}, \citealt{bros2010}, \citealt{bros2013}). The subsequent cooling of the loop plasma is also clearly observed in the evolution of the DEM profiles. For flare kernels, the DEM shows plasma centred around 3\,MK appearing in the impulsive phase of the events. When footpoints as well as loops lie along the LoS, the loop emission dominates but both components can clearly be distinguished in the DEM profiles.
 
 The June 6, 2020 flare shows that STIX allows a detailed, time-resolved study of both thermal plasma and nonthermal electrons in microflares down to GOES class B2. This limit may vary throughout the mission depending on the spacecraft's solar distance.
 The study presented here demonstrates for the first time how detailed multi-instrument studies of solar flares can be performed with STIX. 
 \begin{acknowledgements}
JS and AMV acknowledge the Austrian Science Fund (FWF): I4555-N. AFB and HX are supported by the Swiss National Science Foundation Grant 200021L\_189180 for STIX.
"Solar Orbiter is a space mission of international collaboration between ESA and NASA, operated by ESA."
\end{acknowledgements}

%
%

\bibliographystyle{aa} 
\bibliography{biblio.bib} 

\begin{thebibliography}{42}
\expandafter\ifx\csname natexlab\endcsname\relax\def\natexlab#1{#1}\fi

\bibitem[{{Aschwanden} {et~al.}(2015){Aschwanden}, {Boerner}, {Ryan}, {Caspi},
  {McTiernan}, \& {Warren}}]{aschwanden2015}
{Aschwanden}, M.~J., {Boerner}, P., {Ryan}, D., {et~al.} 2015, \apj, 802, 53

\bibitem[{{Battaglia} {et~al.}(2021){Battaglia}, {Krucker}, {Hurford},
  {Warmuth}, {Veronig}, \& {Dickson}}]{battaglia2021}
{Battaglia}, A., {Krucker}, S., {Hurford}, G., {et~al.} 2021, unpublished

\bibitem[{{Battaglia} {et~al.}(2015){Battaglia}, {Motorina}, \&
  {Kontar}}]{battaglia2015}
{Battaglia}, M., {Motorina}, G., \& {Kontar}, E.~P. 2015, \apj, 815, 73

\bibitem[{{Benz}(2017)}]{benz2017}
{Benz}, A.~O. 2017, Living Reviews in Solar Physics, 14, 2

\bibitem[{{Benz} {et~al.}(2012){Benz}, {Gallagher}, {Veronig}, {Grimm},
  {Sylwester}, {Orleanski}, {Arnold}, {Bednarzik}, {Farnik}, {Hurford},
  {Krucker}, {Limousin}, {Mann}, \& {Vilmer}}]{stix2012}
{Benz}, A.~O., {Gallagher}, P., {Veronig}, A., {et~al.} 2012, IAU Special
  Session, 6, E5.09

\bibitem[{{Berkebile-Stoiser} {et~al.}(2009){Berkebile-Stoiser},
  {G{\"o}m{\"o}ry}, {Veronig}, {Ryb{\'a}k}, \& {S{\"u}tterlin}}]{stoiser2009}
{Berkebile-Stoiser}, S., {G{\"o}m{\"o}ry}, P., {Veronig}, A.~M., {Ryb{\'a}k},
  J., \& {S{\"u}tterlin}, P. 2009, \aap, 505, 811

\bibitem[{{Brosius}(2013)}]{bros2013}
{Brosius}, J.~W. 2013, \apj, 777, 135

\bibitem[{{Brosius} \& {Holman}(2010)}]{bros2010}
{Brosius}, J.~W. \& {Holman}, G.~D. 2010, \apj, 720, 1472

\bibitem[{{Brown}(1971)}]{brown1971}
{Brown}, J.~C. 1971, \solphys, 18, 489

\bibitem[{{Caspi} {et~al.}(2014){Caspi}, {McTiernan}, \& {Warren}}]{caspi2014}
{Caspi}, A., {McTiernan}, J.~M., \& {Warren}, H.~P. 2014, \apjl, 788, L31

\bibitem[{{Cheng} {et~al.}(2012){Cheng}, {Zhang}, {Saar}, \&
  {Ding}}]{cheng2012}
{Cheng}, X., {Zhang}, J., {Saar}, S.~H., \& {Ding}, M.~D. 2012, \apj, 761, 62

\bibitem[{{Christe} {et~al.}(2008){Christe}, {Hannah}, {Krucker}, {McTiernan},
  \& {Lin}}]{christie2008}
{Christe}, S., {Hannah}, I.~G., {Krucker}, S., {McTiernan}, J., \& {Lin}, R.~P.
  2008, \apj, 677, 1385

\bibitem[{{Dennis} \& {Zarro}(1993)}]{dennisZarro1993}
{Dennis}, B.~R. \& {Zarro}, D.~M. 1993, \solphys, 146, 177

\bibitem[{{Dere} {et~al.}(2019){Dere}, {Del Zanna}, {Young}, {Landi}, \&
  {Sutherland}}]{chianti19}
{Dere}, K.~P., {Del Zanna}, G., {Young}, P.~R., {Landi}, E., \& {Sutherland},
  R.~S. 2019, \apjs, 241, 22

\bibitem[{{Dere} {et~al.}(1997){Dere}, {Landi}, {Mason}, {Monsignori Fossi}, \&
  {Young}}]{chianti97}
{Dere}, K.~P., {Landi}, E., {Mason}, H.~E., {Monsignori Fossi}, B.~C., \&
  {Young}, P.~R. 1997, \aaps, 125, 149

\bibitem[{{Fletcher} {et~al.}(2011){Fletcher}, {Dennis}, {Hudson}, {Krucker},
  {Phillips}, {Veronig}, {Battaglia}, {Bone}, {Caspi}, {Chen}, {Gallagher},
  {Grigis}, {Ji}, {Liu}, {Milligan}, \& {Temmer}}]{fletcher2011}
{Fletcher}, L., {Dennis}, B.~R., {Hudson}, H.~S., {et~al.} 2011, \ssr, 159, 19

\bibitem[{{Glesener} {et~al.}(2020){Glesener}, {Krucker}, {Duncan}, {Hannah},
  {Grefenstette}, {Chen}, {Smith}, {White}, \& {Hudson}}]{glesener2020}
{Glesener}, L., {Krucker}, S., {Duncan}, J., {et~al.} 2020, \apjl, 891, L34

\bibitem[{{Hannah} {et~al.}(2008){Hannah}, {Christe}, {Krucker}, {Hurford},
  {Hudson}, \& {Lin}}]{hannah2008}
{Hannah}, I.~G., {Christe}, S., {Krucker}, S., {et~al.} 2008, \apj, 677, 704

\bibitem[{{Hannah} {et~al.}(2011){Hannah}, {Hudson}, {Battaglia}, {Christe},
  {Ka{\v{s}}parov{\'a}}, {Krucker}, {Kundu}, \& {Veronig}}]{hannah2011}
{Hannah}, I.~G., {Hudson}, H.~S., {Battaglia}, M., {et~al.} 2011, \ssr, 159,
  263

\bibitem[{{Hannah} \& {Kontar}(2012)}]{HK2012}
{Hannah}, I.~G. \& {Kontar}, E.~P. 2012, \aap, 539, A146

\bibitem[{{Holman} {et~al.}(2011){Holman}, {Aschwanden}, {Aurass}, {Battaglia},
  {Grigis}, {Kontar}, {Liu}, {Saint-Hilaire}, \& {Zharkova}}]{holman2011}
{Holman}, G.~D., {Aschwanden}, M.~J., {Aurass}, H., {et~al.} 2011, \ssr, 159,
  107

\bibitem[{{Hudson}(1991)}]{hudson1991}
{Hudson}, H.~S. 1991, in Bulletin of the American Astronomical Society,
  Vol.~23, 1064

\bibitem[{{Inglis} \& {Christe}(2014{\natexlab{a}})}]{christe2014}
{Inglis}, A.~R. \& {Christe}, S. 2014{\natexlab{a}}, \apj, 789, 116

\bibitem[{{Inglis} \& {Christe}(2014{\natexlab{b}})}]{inglis2014}
{Inglis}, A.~R. \& {Christe}, S. 2014{\natexlab{b}}, \apj, 789, 116

\bibitem[{{Koskinen} {et~al.}(2017){Koskinen}, {Baker}, {Balogh}, {Gombosi},
  {Veronig}, \& {von Steiger}}]{Koskinen2017}
{Koskinen}, H. E.~J., {Baker}, D.~N., {Balogh}, A., {et~al.} 2017, \ssr, 212,
  1137

\bibitem[{{Krucker} {et~al.}(2020){Krucker}, {Hurford}, {Grimm}, {K{\"o}gl},
  {Gr{\"o}belbauer}, {Etesi}, {Casadei}, {Csillaghy}, {Benz}, {Arnold},
  {Molendini}, {Orleanski}, {Schori}, {Xiao}, {Kuhar}, {Hochmuth}, {Felix},
  {Schramka}, {Marcin}, {Kobler}, {Iseli}, {Dreier}, {Wiehl}, {Kleint},
  {Battaglia}, {Lastufka}, {Sathiapal}, {Lapadula}, {Bednarzik}, {Birrer},
  {Stutz}, {Wild}, {Marone}, {Skup}, {Cichocki}, {Ber}, {Rutkowski}, {Bujwan},
  {Juchnikowski}, {Winkler}, {Darmetko}, {Michalska}, {Seweryn}, {Bia{\l}ek},
  {Osica}, {Sylwester}, {Kowalinski}, {{\'S}cis{\l}owski}, {Siarkowski},
  {St{\k{e}}{\'s}licki}, {Mrozek}, {Podg{\'o}rski}, {Meuris}, {Limousin},
  {Gevin}, {Le Mer}, {Brun}, {Strugarek}, {Vilmer}, {Musset}, {Maksimovi{\'c}},
  {F{\'a}rn{\'\i}k}, {Koz{\'a}{\v{c}}ek}, {Ka{\v{s}}parov{\'a}}, {Mann},
  {{\"O}nel}, {Warmuth}, {Rendtel}, {Anderson}, {Bauer}, {Dionies}, {Paschke},
  {Pl{\"u}schke}, {Woche}, {Schuller}, {Veronig}, {Dickson}, {Gallagher},
  {Maloney}, {Bloomfield}, {Piana}, {Massone}, {Benvenuto}, {Massa},
  {Schwartz}, {Dennis}, {van Beek}, {Rodr{\'\i}guez-Pacheco}, \&
  {Lin}}]{stix2020V2}
{Krucker}, S., {Hurford}, G.~J., {Grimm}, O., {et~al.} 2020, \aap, 642, A15

\bibitem[{{Lemen} {et~al.}(2012){Lemen}, {Title}, {Akin}, {Boerner}, {Chou},
  {Drake}, {Duncan}, {Edwards}, {Friedlaender}, {Heyman}, {Hurlburt}, {Katz},
  {Kushner}, {Levay}, {Lindgren}, {Mathur}, {McFeaters}, {Mitchell}, {Rehse},
  {Schrijver}, {Springer}, {Stern}, {Tarbell}, {Wuelser}, {Wolfson}, {Yanari},
  {Bookbinder}, {Cheimets}, {Caldwell}, {Deluca}, {Gates}, {Golub}, {Park},
  {Podgorski}, {Bush}, {Scherrer}, {Gummin}, {Smith}, {Auker}, {Jerram},
  {Pool}, {Soufli}, {Windt}, {Beardsley}, {Clapp}, {Lang}, \&
  {Waltham}}]{Lemen2012}
{Lemen}, J.~R., {Title}, A.~M., {Akin}, D.~J., {et~al.} 2012, \solphys, 275, 17

\bibitem[{{Lin} \& {Hudson}(1976)}]{linHudson1976}
{Lin}, R.~P. \& {Hudson}, H.~S. 1976, \solphys, 50, 153

\bibitem[{{McTiernan} {et~al.}(2019){McTiernan}, {Caspi}, \&
  {Warren}}]{mctiernan2019}
{McTiernan}, J.~M., {Caspi}, A., \& {Warren}, H.~P. 2019, \apj, 881, 161

\bibitem[{{M{\"u}ller} {et~al.}(2020{\natexlab{a}}){M{\"u}ller}, {St. Cyr},
  {Zouganelis}, {Gilbert}, {Marsden}, {Nieves-Chinchilla}, {Antonucci},
  {Auch{\`e}re}, {Berghmans}, {Horbury}, {Howard}, {Krucker}, {Maksimovic},
  {Owen}, {Rochus}, {Rodriguez-Pacheco}, {Romoli}, {Solanki}, {Bruno},
  {Carlsson}, {Fludra}, {Harra}, {Hassler}, {Livi}, {Louarn}, {Peter},
  {Sch{\"u}hle}, {Teriaca}, {del Toro Iniesta}, {Wimmer-Schweingruber},
  {Marsch}, {Velli}, {De Groof}, {Walsh}, \& {Williams}}]{mueller2020}
{M{\"u}ller}, D., {St. Cyr}, O.~C., {Zouganelis}, I., {et~al.}
  2020{\natexlab{a}}, \aap, 642, A1

\bibitem[{{M{\"u}ller} {et~al.}(2020{\natexlab{b}}){M{\"u}ller}, {St. Cyr},
  {Zouganelis}, {Gilbert}, {Marsden}, {Nieves-Chinchilla}, {Antonucci},
  {Auch{\`e}re}, {Berghmans}, {Horbury}, {Howard}, {Krucker}, {Maksimovic},
  {Owen}, {Rochus}, {Rodriguez-Pacheco}, {Romoli}, {Solanki}, {Bruno},
  {Carlsson}, {Fludra}, {Harra}, {Hassler}, {Livi}, {Louarn}, {Peter},
  {Sch{\"u}hle}, {Teriaca}, {del Toro Iniesta}, {Wimmer-Schweingruber},
  {Marsch}, {Velli}, {De Groof}, {Walsh}, \& {Williams}}]{soloscience2020V2}
{M{\"u}ller}, D., {St. Cyr}, O.~C., {Zouganelis}, I., {et~al.}
  2020{\natexlab{b}}, \aap, 642, A1

\bibitem[{{Neupert}(1968)}]{neupert1968}
{Neupert}, W.~M. 1968, \apjl, 153, L59

\bibitem[{{Podladchikova} {et~al.}(2017){Podladchikova}, {Van der Linden}, \&
  {Veronig}}]{Podladchikova2017}
{Podladchikova}, T., {Van der Linden}, R., \& {Veronig}, A.~M. 2017, \apj, 850,
  81

\bibitem[{{Priest} \& {Forbes}(2002)}]{priestforbes2002}
{Priest}, E.~R. \& {Forbes}, T.~G. 2002, \aapr, 10, 313

\bibitem[{{Schwartz} {et~al.}(2002){Schwartz}, {Csillaghy}, {Tolbert},
  {Hurford}, {McTiernan}, \& {Zarro}}]{schwartz2002}
{Schwartz}, R.~A., {Csillaghy}, A., {Tolbert}, A.~K., {et~al.} 2002, \solphys,
  210, 165

\bibitem[{{Stoiser} {et~al.}(2007){Stoiser}, {Veronig}, {Aurass}, \&
  {Hanslmeier}}]{stoiser2007}
{Stoiser}, S., {Veronig}, A.~M., {Aurass}, H., \& {Hanslmeier}, A. 2007,
  \solphys, 246, 339

\bibitem[{{Vanninathan} {et~al.}(2015){Vanninathan}, {Veronig}, {Dissauer},
  {Madjarska}, {Hannah}, \& {Kontar}}]{vanninathan2015}
{Vanninathan}, K., {Veronig}, A.~M., {Dissauer}, K., {et~al.} 2015, \apj, 812,
  173

\bibitem[{{Veronig} {et~al.}(2002){Veronig}, {Vr{\v{s}}nak}, {Dennis},
  {Temmer}, {Hanslmeier}, \& {Magdaleni{\'c}}}]{Veronig2002}
{Veronig}, A., {Vr{\v{s}}nak}, B., {Dennis}, B.~R., {et~al.} 2002, \aap, 392,
  699

\bibitem[{{Veronig} {et~al.}(2005){Veronig}, {Brown}, {Dennis}, {Schwartz},
  {Sui}, \& {Tolbert}}]{veronig2005}
{Veronig}, A.~M., {Brown}, J.~C., {Dennis}, B.~R., {et~al.} 2005, \apj, 621,
  482

\bibitem[{{Warmuth} \& {Mann}(2020)}]{warmuth2020}
{Warmuth}, A. \& {Mann}, G. 2020, \aap, 644, A172

\bibitem[{{White} {et~al.}(2005){White}, {Thomas}, \& {Schwartz}}]{goesEMT2005}
{White}, S.~M., {Thomas}, R.~J., \& {Schwartz}, R.~A. 2005, \solphys, 227, 231

\bibitem[{{Wright} {et~al.}(2017){Wright}, {Hannah}, {Grefenstette},
  {Glesener}, {Krucker}, {Hudson}, {Smith}, {Marsh}, {White}, \&
  {Kuhar}}]{wright2017}
{Wright}, P.~J., {Hannah}, I.~G., {Grefenstette}, B.~W., {et~al.} 2017, \apj,
  844, 132

\end{thebibliography}

\end{document}